\documentclass[twocolumn,amsmath,amssymb,superscriptaddress,nofootinbib]{revtex4-2}
\usepackage{graphicx,amsmath,relsize,epstopdf,color,mathtools,bm,newtxtext,newtxmath,rotating,dsfont,stackrel}
\usepackage[hyphenbreaks]{breakurl}
\usepackage[colorlinks=true,linkcolor=blue,citecolor=blue,urlcolor =blue]{hyperref}

\newcommand{\eq}[1]{\begin{equation}\begin{aligned}#1\end{aligned}\end{equation}}
\newcommand{\iu}{\text{i}}
\newcommand{\eu}{\text{e}}
\newcommand{\ha}{\hat{a}}
\newcommand{\had}{\hat{a}^\dagger\vphantom{a}}
\newcommand{\hb}{\hat{b}}
\newcommand{\hbd}{\hat{b}^\dagger\vphantom{a}}
\newcommand{\hoa}{\hat{O}_a}
\newcommand{\hoad}{\hat{O}_a^\dagger\vphantom{a}}
\newcommand{\hob}{\hat{O}_b}
\newcommand{\hobd}{\hat{O}_b^\dagger\vphantom{a}}
\newcommand{\ket}[1]{\left\lvert#1\right\rangle}

\newcommand{\bra}[1]{\left\langle#1\right\rvert}

\newcommand{\braket}[2]{\left.\left\langle#1\right\rvert#2\right\rangle}
\newcommand{\Tr}{\mathop{\mathrm{Tr}} \nolimits}

\newcommand{\Var}{\mathop{\mathrm{Var}}\nolimits}

\usepackage[table,xcdraw]{xcolor}
\newcommand{\expct}[1]{\left\langle#1\right\rangle}
\newcommand{\RE}{\mathop{\mathrm{Re}} \nolimits}

\begin{document}

\title{How squeezed states both maximize and minimize the same notion of quantumness}
\author{Aaron Z. Goldberg}
\affiliation{National Research Council of Canada, 100 Sussex Drive, Ottawa, Ontario K1A 0R6, Canada}
\affiliation{Department of Physics, University of Toronto, 60 St. George Street, Toronto, Ontario, M5S 1A7, Canada}

\author{Khabat Heshami}
\affiliation{National Research Council of Canada, 100 Sussex Drive, Ottawa, Ontario K1A 0R6, Canada}
\affiliation{Department of Physics, University of Ottawa, Advanced Research Complex,
	25 Templeton Street, Ottawa ON Canada, K1N 6N5}
\affiliation{Institute for Quantum Science and Technology, Department of Physics and Astronomy, University of Calgary, Alberta T2N 1N4, Canada}

\begin{abstract}
Beam splitters are routinely used for generating entanglement between modes in the optical and microwave domains, requiring input states that are not convex combinations of coherent states. This leads to the ability to generate entanglement at a beam splitter as a notion of quantumness. 
A similar, yet distinct, notion of quantumness is the amount of entanglement generated by two-mode squeezers (i.e., four-wave mixers).
We show that squeezed-vacuum states, paradoxically, both minimize and maximize these notions of quantumness, with the crucial resolution of the paradox hinging upon the relative phases between the input states and the devices. 
Our notion of quantumness is intrinsically related to eigenvalue equations involving creation and annihilation operators, governed by a set of inequalities that leads to generalized cat and squeezed-vacuum states. 
%
\end{abstract}

\maketitle

\section{Introduction}
Many fundamental differences between quantum and classical mechanics involve quantum entanglement \cite{EPR1935,Bell1964,Bennettetal1993,Jenneweinetal2000,Giovannettietal2006}. One of the simplest methods for generating entanglement is by impinging ``nonclassical'' states of light on a beam splitter \cite{Tanetal1991,Sanders1992,HuangAgarwal1994,KimSanders1996,Paris1999,Kimetal2002,Xiangbin2002,Wolfetal2003,Asbothetal2005,Tahiraetal2009,Pianietal2011,Jiangetal2013,Killoranetal2014,VogelSperling2014,Geetal2015,Streltsovetal2015,GholipourShahandeh2016,Maetal2016,GoldbergJames2018,Fuetal2020}, where ``classical'' implies a convex mixture of coherent states \cite{Sudarshan1963,Glauber1963,BachLuxmannEllinghaus1986,Sperling2016}. This leads us to define a notion of ``quantumness'' as the ability of an input state to generate entangled outputs at a beam splitter. Here, we answer: What quantum states maximize and minimize this notion?

Our framework applies to other elements that are routinely used for generating entanglement in the optical and microwave domains, such as four-wave mixers that generate two-mode squeezing. Moreover, it extends to other physical scenarios such as cavity optomechanics \cite{Purdyetal2013,Aspelmeyeretal2014}, where the distinction between a beam splitter and a two-mode squeezer simply depends on the relationship between the laser-cavity detuning and the mechanical frequency. We thus address the more general question: What quantum states maximize and minimize the amount of entanglement generated at an arbitrary element?

It is well known that, when one of the two states input to a beam splitter is the vacuum state, any state in the second mode that is not a convex combination of coherent states will generate entanglement. This led to the notion of entanglement potential \cite{Asbothetal2005}, which measures the amount of entanglement generated by a single nonvacuum state input to a beam splitter. However, when both input modes are not in their vacuum states, there are nonclassical, non-Gaussian, and mixed states that generate no entanglement at beam splitters \cite{HuangAgarwal1994,KimSanders1996,Jiangetal2013,GoldbergJames2018}.  A further notion of quantumness is thus the amount of entanglement generated by a general \textit{two-mode} state impinging on a beam splitter. In light of recent work \cite{Goldbergetal2020extremal}, we seek states that extremize this notion, in order to constrain the possibilities allowed by quantum theory.

Discussions regarding the entanglement generated by beam splitters inevitably raise questions about the nature of mode entanglement versus particle entanglement. Mathematically, all entanglement is created equally while, physically, only certain types of entanglement may be useful \cite{Ghirardietal2002,Zanardietal2004,Barnumetal2004,HarshmanWickramesekara2007,Horodeckietal2009,Karimi1Boyd2015}. This has been extensively explored in the literature \cite{PaskauskasYou2001,Lietal2001,Eckertetal2002,WisemanVaccaro2003,Ghirardietal2004,WangSanders2005,Ieminietal2014,Reuschetal2015,Benattietal2017,Morrisetal2020,Benattietal2020}, especially some time ago in the context of single-photon entanglement with the vacuum \cite{Tanetal1991,Gerry1996,Bjorketal2001,Lombardietal2002,Leeetal2003,Hessmoetal2004,Babichevetal2004,vanEnk2005,PawlowskiCzachor2006,Drezet2006,vanEnk2006reply,DiFidioVogel2009,Salartetal2010,Leuchs2016,Dasetal2021arxiv}; our current focus is not to readdress those questions but to simply ask how much entanglement a state can generate while reserving judgement about the usefulness of such entanglement.

Further, we can circumvent distinctions between various types of entanglement by considering other elements known to generate useful entanglement, such as two-mode squeezers. Coherent states again minimize the amount of entanglement generated at two-mode squeezers, leading to a definition of quantumness as the amount of entanglement generated therewith, which shows the reach of our notion of quantumness beyond particular types of entanglement.

Our resulting most and least quantum states satisfy the eigenvalue equation
\eq{
\ha\ket{\Psi}=\eta\had\ket{\Psi},
\label{eq:creation annihilation eigenvalue}
} where $\ha$ is the annihilation operator for a bosonic mode and $\eta$ is a complex-valued constant. Equation \eqref{eq:creation annihilation eigenvalue} is one of the defining relationships of squeezed-vacuum states \cite{Kennard1927,MillerMishkin1966,Stoler1970,Lu1971,Dodonov2002}. It turns out that the relationship between the phases of the squeezed states input to an optical element and the phase applied by the optical element governs the transition from generating the most entanglement to generating the least entanglement, which we find to broadly hold for a variety of optical elements. Here, and in the rest of this work, we mainly refer to optical elements, but our results broadly extend to the microwave domain and beyond.

In exploring this notion of quantumness, we look for states satisfying the generalized eigenvalue equation
\eq{
\ha^k\ket{\Psi}=\eta\had^l\ket{\Psi}
\label{eq:creation annihilation eigenvalue kl}
} for any pair of integers $k$ and $l$ with $k\geq l$, which was very recently explored in the case of $k=l$ and $\eta=1$ \cite{PereverzevBittner2021}. This leads us to a set of generalized squeezed-vacuum states, whose peculiar phase-space properties we investigate in detail. We also detail how our inequalities involving creation and annihilation operators lead to the generalizations of cat states \cite{Dodonovetal1974} to compass states \cite{Zurek2001}. Finally, we discuss the usefulness of these generalized states for tasks such as metrology and the ease with which they may be generated using nonlinear optical devices.

\section{Using beam splitters to generate entanglement}
We begin by considering two orthogonal modes annihilated by bosonic operators $\ha$ and $\hb$. These modes can be, for example, two spatial or two polarization modes of light. A generic beam splitter can be represented by an SU(2) operator $\hat{U}$ that enacts \cite{LuisSanchezSoto1995}
\eq{
\hat{U}^\dagger\begin{pmatrix}\ha\\\hb
\end{pmatrix}\hat{U}=
\begin{pmatrix}
\eu^{-\iu\frac{\phi+\psi}{2}}\cos\frac{\theta}{2}
&
-\eu^{-\iu\frac{\phi-\psi}{2}}\sin\frac{\theta}{2}\\
\eu^{\iu\frac{\phi-\psi}{2}}\sin\frac{\theta}{2}
&
\eu^{\iu\frac{\phi+\psi}{2}}\cos\frac{\theta}{2}
\end{pmatrix}
\begin{pmatrix}\ha\\\hb
\end{pmatrix}.
\label{eq:beam splitter matrix}
}
We define separable pure states as those that can be decomposed via the tensor product
\eq{
\ket{\Phi_{\mathrm{sep}}^{(a,b)}}=\ket{\Psi^{(a)}}_a\otimes \ket{\Psi^{(b)}}_b
} and entangled pure states as those that cannot. We will discard the tensor product symbol $\otimes$ and mode subscripts in what follows unless required for clarity.
These definitions make it clear that the two-mode coherent states
\eq{
\ket{\alpha}\ket{\beta}=
\eu^{\alpha \had-\alpha^*\ha}\ket{0} \eu^{\beta \hbd-\beta^*\hb}\ket{0}
} remain separable following a beam-splitter transformation:
\eq{
\hat{U}\ket{\alpha}\ket{\beta}= &\ket{\eu^{-\iu\frac{\phi+\psi}{2}}\cos\frac{\theta}{2}\alpha+\eu^{\iu\frac{\phi-\psi}{2}}\sin\frac{\theta}{2}\beta}
\\&\otimes
\ket{-\eu^{-\iu\frac{\phi-\psi}{2}}\sin\frac{\theta}{2}\alpha+\eu^{\iu\frac{\phi+\psi}{2}}\cos\frac{\theta}{2}\beta}.
}

Since separable mixed states are defined as convex combinations of separable pure states, and entangled mixed states as those without such a decomposition, convex combinations of two-mode coherent states, which are the only states whose Glauber-Sudarshan $P$-functions are positive everywhere \cite{Sudarshan1963,Glauber1963}, are seen to generate no entanglement at beam splitters. This yields a necessary but not sufficient condition for generating entanglement: the input states must not be convex combinations of coherent states. That the condition is not sufficient can be seen using squeezed states of light: the state
\eq{
\ket{\Phi_{\mathrm{sep}}^{(a,b)}}=\hat{S}_a(r_a,\varphi_a)\ket{0}\hat{S}_b(r_a,\varphi_a+2\psi)\ket{0}
,
\label{eq:two modes squeezed no ent}
} where we have defined the single-mode squeezing operator 
\eq{
\hat{S}_a(r_a,\varphi_a)=\exp\left({r_a\frac{\eu^{-\iu\varphi_a}\ha^2-\eu^{\iu\varphi_a}\had^2}{2}}\right)
\label{eq:single-mode squeezing operator}
} and similarly for mode $b$, will remain separable after undergoing the beam-splitter transformation $\hat{U}$.
We note here the crucial dependence of the input states' relative phase on the beam splitter phase $\psi$.
All other separable pure states will generate entanglement via $\hat{U}$.

There are even more separable two-mode mixed states that do not generate entanglement via $\hat{U}$ \cite{GoldbergJames2018}. This significantly differs from the single-mode case, in which $P$-function negativity is necessary and sufficient for characterizing the present notion of quantumness, and motivates a study of the potential for two-mode states to generate entanglement using beam splitters.

\section{Extremizing the entanglement generated by optical devices}
\subsection{Weak beam splitters}
To search for the states with the most and least quantumness, we first seek states that are the most and least able to generate entanglement at weakly-reflecting (or, equivalently, weakly-transmitting) beam splitters.
These states require the minimal amount of assistance from beam splitters in order to demonstrate their quantumness. Weakly-reflecting beam splitters are represented by transformations of the form of Eq. \eqref{eq:beam splitter matrix} with small $\theta$ and have the same entanglement properties as weakly-transmitting beam splitters with small $\pi-\theta$.

We quantify the amount of entanglement generated by a transformation $\hat{U}$ using the linear entropy
\eq{
H\left(\Phi_{\mathrm{sep}}^{(a,b)}\right)=1-\Tr\left[\left(\Tr_a \hat{\rho}\right)^2\right]=1-\Tr\left[\left(\Tr_b \hat{\rho}\right)^2\right],
} where \eq{
\hat{\rho}=\hat{U}\ket{\Phi_{\mathrm{sep}}^{(a,b)}}\bra{\Phi_{\mathrm{sep}}^{(a,b)}}\hat{U}^\dagger
} and $\Tr_c$ is the partial trace with respect to mode $c\in(a,b)$.  A linear entropy $H\left(\Phi_{\mathrm{sep}}^{(a,b)}\right)=0$ implies that the initial state $\ket{\Phi_{\mathrm{sep}}^{(a,b)}}$ generates no entanglement via the beam-splitter transformation $\hat{U}$; $H$ monotonically increases to 1 with increasing entanglement of the output state.

We calculate the linear entropy for this transformation and generalize it to other transformations in Appendix \ref{app:linear entropy general unitary}. The final result, to order $\mathcal{O}(\theta^2)$, is:
\eq{
H\left(\Phi_{\mathrm{sep}}^{(a,b)}\right)&=\theta^2\left\{\frac{A+B}{2}+AB-\RE\left[\eu^{-2\iu\psi}\Var\left(\had\right)\Var\left(\hb\right)\right]\right\}.
\label{eq:linear entropy second order}
} Here, we have defined
the terms
\eq{
A&=
\expct{\had\ha}-\left|\expct{\ha}\right|^2, \qquad
B=\expct{\hbd\hb}-\left|\expct{\hb}\right|^2,
} and $\Var(\hat{X})=\expct{\hat{X}^2}-\expct{\hat{X}}^2$, where the expectation values defining functions of $\ha$ and $\had$ can be taken with respect to $\ket{\Phi_{\mathrm{sep}}^{(a,b)}}$ or $\ket{\Psi^{(a)}}$
and those defining functions of $\hb$ and $\hbd$ can be taken with respect to $\ket{\Phi_{\mathrm{sep}}^{(a,b)}}$ or $\ket{\Psi^{(b)}}$.

The amount of entanglement generated by a separable state increases with $\theta$; this is why we choose small $\theta$, to isolate the entanglement-generating properties of the input states from those attributed to the beam splitter itself. Since the terms $A$ and $B$ are always positive, the amount of entanglement generated seems to increase with decreasing magnitudes of $\expct{\ha}$ and $\expct{\hb}$. Indeed, for a fixed total input energy, which is proportional to $N=\expct{\had\ha+\hbd\hb}$ and remains unchanged by $\hat{U}$, the amount of entanglement generated does increase as the expectation values of $\ha$ and $\hb$ decrease in magnitude. Coherent states satisfy the eigenvalue equation
$
\ha\ket{\alpha}=\alpha\ket{\alpha}$ \cite{Klauder1960}, from which it is apparent that $H(\alpha,\beta)=0$. In this sense we can say that the two-mode coherent states have the least quantumness.

The pair of single-mode squeezed states discussed earlier in Eq. \eqref{eq:two modes squeezed no ent} also minimizes this notion of quantumness. This is because such states satisfy $\expct{\ha}=0$, $\expct{\hb}=0$, and
\eq{
f(A,B)\equiv \frac{A+B}{2}+AB=\RE\left[\eu^{-2\iu\psi}\Var\left(\had\right)\Var\left(\hb\right)\right].
} A peculiar, particular arrangement between the phases of two squeezed states with equal squeezing strength and the relative phase imparted by the beam splitter prevents these supposedly nonclassical states from generating entanglement.

We can similarly find the states that generate the most entanglement for a given total input energy. To begin, we notice that $A,B\geq 0$, with equality if and only if modes $a$ and $b$ house coherent states, implying that the function $f(A,B)$ obeys the inequality
\eq{
f(A,B)\geq 0,
} with equality if and only if both modes $a$ and $b$ are occupied by coherent states. The function $f(A,B)$ is maximized by $\expct{\ha}=\expct{\hb}=0$. Recalling that we have fixed the total energy, the maximum value of $f(A,B)$ only depends on the initial energy difference between the two modes, proportional to $\expct{\had\ha-\hbd\hb}$, achieving the maximum value when $A=B=N/2$:
\eq{
f_{\mathrm{max}}(A,B; N)=\frac{N}{2}\left(\frac{N}{2}+1\right).
}

Next, we aim to maximize
$-\RE\left[\eu^{-2\iu\psi}\Var\left(\had\right)\Var\left(\hb\right)\right]$. Since the variances in question deal only with the creation operator from one mode paired with the annihilation operator from the other, changing the relative phase between the input states will change the absolute phase of $\Var\left(\had\right)\Var\left(\hb\right)$. We can thus independently find the optimal phase and find the maximum value of
\eq{
g\left(\Phi_{\mathrm{sep}}^{(a,b)};N\right)=\left|\Var\left(\had\right)\Var\left(\hb\right)\right|.
} By again setting $\expct{\ha}=\expct{\hb}=0$, we find
\eq{
g\left(\Phi_{\mathrm{sep}}^{(a,b)};N\right)=\left|\expct{\ha^2}\right|\left|\expct{\hb^2}\right|.
} We thus seek to maximize $\left|\expct{\ha^2}\right|$ and $\left|\expct{\hb^2}\right|$ while hoping this to accord with the requirements for $f_\mathrm{max}$.

One avenue that does not fully solve the problem is to consider the Cauchy-Schwarz inequality
\eq{
\left|\expct{\ha^2}\right|\leq\sqrt{\expct{\had\had\ha\ha}}=\sqrt{\Var\left(\had\ha\right)+\expct{\had\ha}^2-\expct{\had\ha}}.
\label{eq:CS-cat}
} This inequality is saturated by states satisfying the eigenvalue equation
\eq{
\ha^2\ket{\Psi^{(a)}}\propto \ket{\Psi^{(a)}},
} which is achieved by the cat or Yurke-Stoler states \cite{Dodonovetal1974,YurkeStoler1986}
\eq{
\ket{\Psi^{(a)}}=\ket{\mathrm{cat}}\equiv \frac{\ket{\alpha}+\eu^{\iu\varphi_a}\ket{-\alpha}}{\mathcal{N}},
} where $\mathcal{N}$ is some normalization constant that depends on $\alpha$ and the relative phase $\varphi_a$. However, these states do not maximize the value of $\left|\expct{\ha^2}\right|$ for a fixed $\expct{\had\ha}$, only maximizing the former for certain values of the latter and $\Var\left(\had\ha\right)$.

We can instead solve the problem by considering another Cauchy-Schwarz inequality:
\eq{
\left|\expct{\ha^2}\right|\leq \sqrt{\expct{\had\ha}\expct{\ha\had}}=\sqrt{\expct{\had\ha}\left(\expct{\had\ha}+1\right)}.
\label{eq:CS-squeeze}
} The upper bound of the inequality depends only on the energy of the state $\expct{\had\ha}$ and is saturated by states satisfying the eigenvalue equation mentioned earlier in Eq. \eqref{eq:creation annihilation eigenvalue}. Since that eigenvalue equation defines squeezed states, the latter saturate the bound for a fixed energy. One can then calculate using properties of squeezed states that
\eq{
g_{\mathrm{max}}\left(\Phi_{\mathrm{sep}}^{(a,b)};N\right)=\frac{N}{2}\left(\frac{N}{2}+1\right).
}

Finally, we adjust the phase of $\Var\left(\had\right)\Var\left(\hb\right)$ by adjusting the relative phases $\varphi_a$ and $\varphi_b$ of the squeezing operators for the two input modes. Choosing $\exp\left[\iu\left(\varphi_b-\varphi_a-2\psi\right)\right]=-1$, we see that the required relative phase is $\pi$ different from that required for the states generating no entanglement. We arrive at the conclusion that the pair of single-mode squeezed states
\eq{
\ket{\Phi_{\mathrm{sep}}^{(a,b)}}=\hat{S}_a(r_a,\varphi_a)\ket{0}\hat{S}_b(r_a,\varphi_a+2\psi+\pi)\ket{0}
,
\label{eq:two modes squeezed max ent}
} generates the maximum amount of entanglement at a beam splitter for a fixed input energy $N$.

Equations \eqref{eq:two modes squeezed no ent} and \eqref{eq:two modes squeezed max ent} are very similar, differing only by the extra relative phase of $\pi$ between the two input squeezed states. A given pair of squeezed states with equal squeezing strength $r_a=r_b$ will thus generate the most amount of entanglement for beam splitters with certain phases and the least amount of entanglement for others! In a phase-space picture, this means that the angle required to rotate the phase-space distribution of one input squeezed state such that it completely overlaps with the phase-space distribution of the other input state determines how much entanglement the pair of states will generate. When this rotation angle is $2\psi$, they will generate no entanglement and, when this rotation angle is increase to $2\psi+\pi$, the amount of entanglement generated will monotonically increase to its maximum value,
\eq{
\frac{H_\mathrm{max}\left(\Phi_{\mathrm{sep}}^{(a,b)};N\right)}{\theta^2}=N\left(\frac{N}{2}+1\right).
}

\subsection{Weak two-mode squeezers}
All of the above calculations for entanglement generation  at beam splitters can similarly be performed for entanglement generation by other optical devices. As shown in Appendix \ref{app:linear entropy general unitary}, we can consider optical devices such as two-mode squeezers represented by the unitary operators \eq{{\hat{\mathcal{U}}}=\exp\left(r\frac{\eu^{-\iu\psi}\ha\hb-\eu^{\iu\psi}\had\hbd}{2}\right).}
A separable state input to such a two-mode squeezer with squeezing strength $r$ and phase $\psi$ will yield the linear entropy up until $\mathcal{O}(r^2)$
\eq{
	\frac{2H_{{\hat{\mathcal{U}}}}(\Phi)}{r^2}\approx2AB+A+B+1-2\RE\left[\eu^{-2\iu\psi}\Var\left(\ha\right)\Var\left(\hb\right)\right].
	\label{eq:linear entropy two mode squeeze}
} The extra constant term $1$ and the replacement of $\Var\left(\had\right)$ in Eq. \eqref{eq:linear entropy second order} with $\Var\left(\ha\right)$ here showcase the differences between how beam splitters and two-mode squeezers generate entanglement. 
All of the previous calculations are still relevant, with two-mode coherent states generating the \textit{minimum} amount of entanglement, two single-mode squeezed states with equal squeezing strengths and a particular phase relationship generating the \textit{minimum} amount of entanglement, and 
two single-mode squeezed states with equal squeezing strengths and another particular phase relationship generating the \textit{maximum} amount of entanglement. We again see that squeezed states both minimize and maximize the same notion of quantumness.

Our observation that entanglement generation crucially depends on the phases of the input states again holds.  Here, in contrast to the beam splitter condition, we have a condition for the sum of the phases of the input squeezed states: minimal entanglement is generated when the phase sum satisfies $\varphi_a+\varphi_b=2\psi$ and maximal entanglement is generated when it obeys $\varphi_a+\varphi_b=2\psi+\pi$. We observe the general principle that changing the phase relationship between the input squeezed states and the optical element can tune the entanglement generated from its minimum to its maximum value. {These results expand on the crucial phase relationship between the squeezing of a signal field and of a local oscillator mode when doing homodyne measurement \cite{KimSanders1996}.} Of further note, the extra constant term $1$ in Eq. \eqref{eq:linear entropy two mode squeeze} implies that all states will generate \textit{some} entanglement at a two-mode squeezer, in contrast to a beam splitter that allows certain states to generate \textit{zero} entangement.

\subsection{General observations}
In all cases, only the least quantum states, viz., the two-mode coherent states, do not have their entanglement generated depend on the phase of the optical element in question. Squeezed states, which are the most quantum states according this notion because they generate the most entanglement at these optical elements, are the opposite: their entanglement generated is the most sensitive to the phase of the optical element. This could be used as an alternative notion of quantumness: the most (least) quantum states are those whose entanglement generated at a beam splitter or two-mode squeezer is the most (least) sensitive to the phase of said optical element.

How useful is this entanglement that is generated? This entanglement can be generated even with lossy beam splitters via diffraction \cite{GoldbergJames2019,Sadanaetal2019}. It is clear that not all entangled states can be generated by impinging a separable state onto a beam splitter \cite{MoyanoFernandezGarciaEscartin2017}, while the ability to make post-selective measurements would add significant advantages \cite{Knilletal2001}. Some of the entangled states that can be created by beam splitters are useful, such as squeezed-state inputs in the context of providing quantum enhancements to phase sensing \cite{Caves1981}, while other input states, such as the Fock states $|N\rangle|0\rangle$, provide no quantum enhancements in the context of phase sensing.
When the optical element in question performs two-mode squeezing, the entanglement generated is considered to be much more useful, with applications such as SU(1,1) interferometry \cite{Yurkeetal1986,Caves2020}, quantum barcode reading \cite{Pirandola2011}, quantum-enhanced radar \cite{Changetal2019} and quantum-enhanced spectroscopy \cite{Shietal2020}. In fact, since single-mode squeezing facilitates photonic quantum computing
\cite{Bourassaetal2021blueprintscalable}, it is likely that so too would the entanglement generated by two-mode squeezing. Our investigation of the states generating the most and least entanglement may thus be considered in an device-agnostic manner, with particular implementations being more useful for particular tasks.

\section{Extensions of quantumness indicator lead to generalized cat and squeezed states}
The inequalities in Eqs. \eqref{eq:CS-cat} and \eqref{eq:CS-squeeze} have generalizations that are important for investigating quantumness. 
What states maximize the magnitude of the expectation value of arbitrary powers of annihilation operators $\left|\expct{\ha^n}\right|^2$?
Each type of inequality leads to different results that we study in turn.
\subsection{Generalized cat states}
\begin{figure*}
    \centering
    \includegraphics[trim={130 0 90 0},clip,width=\textwidth]{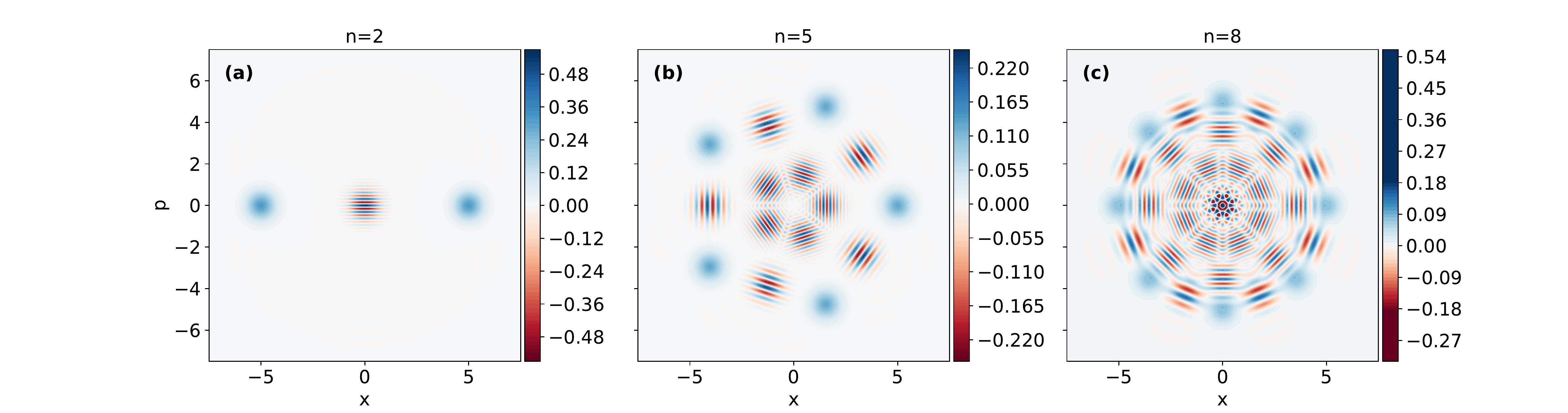}
    \caption{Wigner quasiprobability distributions for higher-order cat states with $\alpha=5$ and $n=2$, $5$, and $8$. These states saturate the inequality of Eq. \eqref{eq:CS-cat-high} and can be considered to maximize that form of quantumness.
    The $x$ and $y$ axes correspond to the dimensionless $x$ and $p$ quadratures of a harmonic oscillator, such as an electromagnetic field, in units where $\left[\hat{x},\hat{p}\right]=\iu$. The Wigner functions are large near the values $\alpha \eu^{2\pi\iu k/n}$ and display their rotational symmetries. They also display their oscillatory structure wherein different coherent states interfere and the Wigner functions take negative values. All figures were created using QuTiP \cite{Johanssonetal2012,Johanssonetal2013}.}
    \label{fig:high cats}
\end{figure*}

We first consider extending Eq. \eqref{eq:CS-cat} to
\eq{
\left|\expct{\ha^n}\right|^2\leq \expct{\had^n\ha^n}.
\label{eq:CS-cat-high}
} This inequality is saturated by the higher-order cat states
\eq{
\ket{\mathrm{high\,cat}}\propto \sum_{k=1}^n \ket{\alpha \eu^{\frac{2\pi\iu k}{n}}}.
} These states are unchanged by rotations of $2\pi/n$ in phase space, as exemplified by the Wigner quasiprobability distribution
\eq{
W_{\Psi}(x,p)=\frac{1}{2\pi}\int_{-\infty}^\infty d\tau\,\eu^{-\iu p\tau}\braket{x-\frac{\tau}{2}}{\Psi}\braket{\Psi \vphantom{\frac{\tau}{2}}}{x+\frac{\tau}{2}}
;}
we display some such distributions in Fig. \ref{fig:high cats}. 

Peculiarly, there are some optical devices with which cat states generate the \textit{least} entanglement. Suppose we have a highly nonlinear optical element represented by the unitary operator ${\hat{\mathcal{U}}}=\exp\left(r\frac{\eu^{-\iu\psi}\had^m\hb^n-\eu^{\iu\psi}\ha^m\hbd^n}{2}\right)$ for integers $m$ and $n$ greater than unity. The general calculations in Appendix \ref{app:linear entropy general unitary} dictate that the simultaneous eigenstates of $\ha^m$ and $\hb^n$ will generate the least entanglement with this highly nonlinear device. As such, two high-order cat states
\eq{
	\ket{\mathrm{two\,high\,cats}}\propto\sum_{k=1}^m \ket{\alpha \eu^{\frac{2\pi\iu k}{m}}}\sum_{l=1}^n \ket{\beta \eu^{\frac{2\pi\iu l}{n}}},
} which by most standards would be considered highly quantum, combined with a highly nonlinear optical element, which would readily generate entanglement with most input states, somehow lead to no entanglement being generated. This, at the very least, shows how intricate the ability to generate entanglement is.

The generalized cat states, also known as compass states \cite{Zurek2001}, are useful for metrological tasks such as sensing displacements in arbitrary directions \cite{Dalvitetal2006,Goldbergetal2020extremal,Akhtaretal2021}, which has been used as another indicator of quantumness \cite{Goldbergetal2020extremal}, among other quantum information tasks including quantum error correction \cite{Sanders1992,Cochraneetal1999,Mirrahimietal2014,BergmannvanLoock2016,Grimsmoetal2020}. As well, there exist numerous proposols for their experimental generation \cite{Leeetal1994,vanEnk2003,Dalvitetal2006,ChoudhuryPanigrahi2011} and they have indeed been generated up until $n=4$ \cite{Vlastakisetal2013,Praxmeyeretal2015}.

Cat states are known to be highly sensitive to their preparation and measurement procedures \cite{Raeisietal2011,Wangetal2013}. For example, their generation requires precise control over the phase of the optical element in question \cite{Wangetal2013}, which may be intimately connected to the sensitivity of the amount of entanglement generated by optical elements to the phases of the input states. This extreme sensitivity to phase relationships seems to be a hallmark of quantum states that is certainly less pronounced for less quantum states such as two-mode coherent states.

\subsection{Generalized squeezed states}
The upper bound of the inequality in Eq. \eqref{eq:CS-cat-high} depends on more than just $\expct{\had\ha}$. This means that saturating the inequality does not guarantee the maximization of $\left|\expct{\ha^n}\right|$. We are prompted to consider other inequalities for maximizing $\left|\expct{\ha^n}\right|$.
A number of generalizations of Eq. \eqref{eq:CS-squeeze} are given by
\eq
{
\left|\expct{\ha^n}\right|^2\leq \expct{\had^k\ha^k}\expct{\ha^{l}\had^{l}},\quad k+l=n,
\label{eq:CS-squeeze-high}
}where we allow any pair of integers $k$ and $l$ satisfying $k\geq l$ (i.e., $k\geq n/2$). These inequalities are saturated by states with the form given in Eq. \eqref{eq:creation annihilation eigenvalue kl} and include the higher-order cat states in the case of $l=0$. 

What are the states satisfying the eigenvalue equation of Eq. \eqref{eq:creation annihilation eigenvalue kl}, which saturate the inequality of Eq. \eqref{eq:CS-squeeze-high}?
The eigenvalue equation determines a recursion relation for the set of coefficients $\left\{\Psi_m\right\}$ in the expansion of the state in the photon-number basis:
\eq{\ket{\Psi}=\sum_m\Psi_m\ket{m}.} The recursion relation is:
\eq{
\Psi_{m+k}\sqrt{\frac{(m+k)!}{m!}}&=\eta \Psi_{m-l}\sqrt{\frac{m!}{(m-l)!}}
.
\label{eq:recursion relation}
} By the ratio test, the series converges for all $k> l$ and all $\eta$, as well as for $k=l$ with $\left|\eta\right|\leq 1$. There are $l$ independent solutions for a given value of $\eta$, each determined by specifying which of the coefficients $\left\{\Psi_0,\Psi_1,\cdots,\Psi_{l-1}\right\}$ should be nonzero. Superpositions of these states will also satisfy the eigenvalue equation in Eq. \eqref{eq:creation annihilation eigenvalue kl}, providing additional freedom for finding the states that maximize the upper bound in the inequality of Eq. \eqref{eq:CS-squeeze-high}.

\begin{figure*}
    \centering
    \includegraphics[trim={100 150 60 150},clip,width=0.75\textwidth]{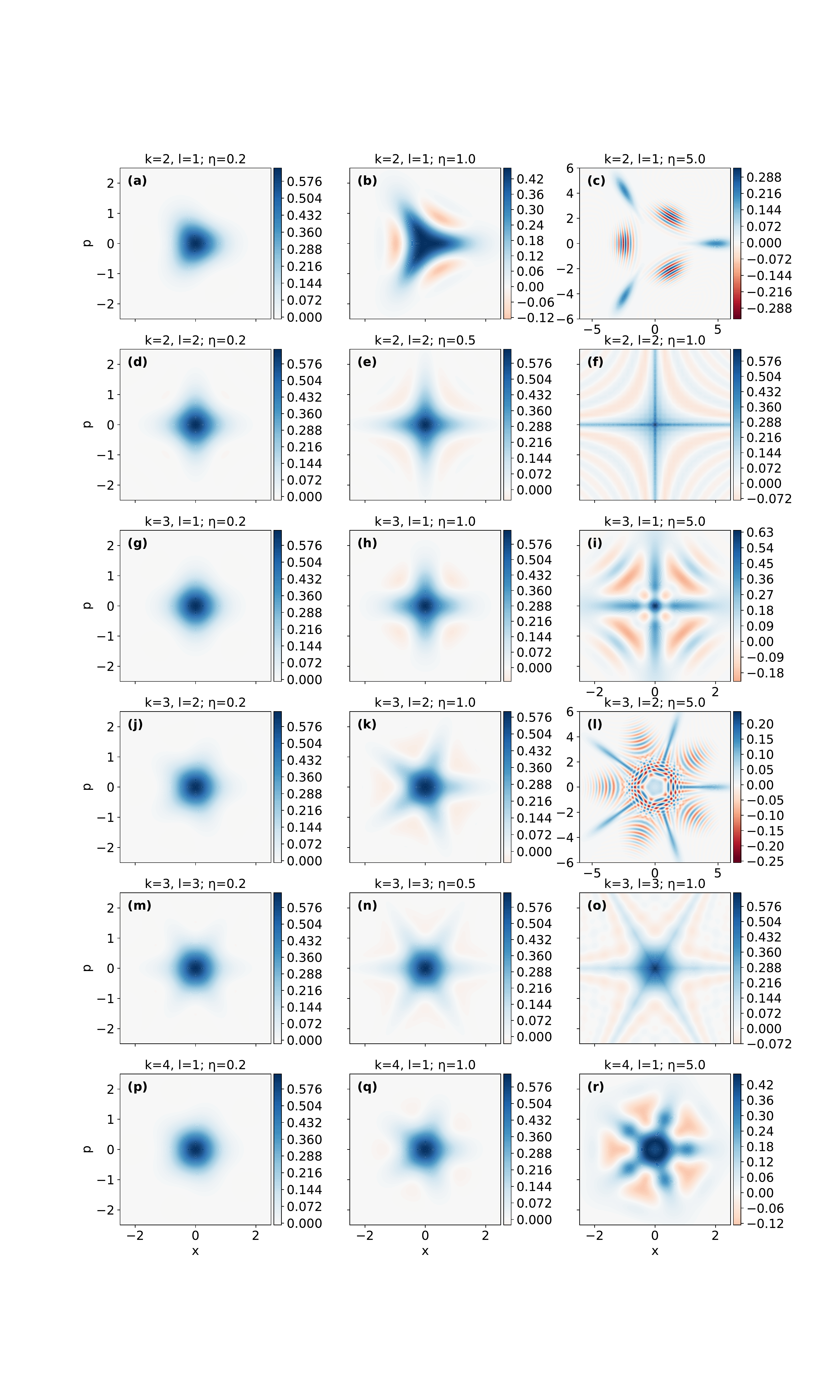}
    \caption{Wigner quasiprobability distributions for the generalized states satisfying the eigenvalue equation of Eq. \eqref{eq:creation annihilation eigenvalue kl} and saturating the inequality of Eq. \eqref{eq:CS-squeeze-high} for values of $k$ and $l$ ranging from 1 to 4 and various values of $\eta$ ($0.2$, $1.0$, and $5.0$ for $k>l$ and $0.2$, $0.5$, and $1.0$ for $k=l$). The states display squeezing from $n=k+l$ directions, with the strength of the squeezing increasing with $|\eta|$. Changing the phase of $\eta$ will rigidly rotate these distributions about the origin. Some negative values of the Wigner function are visible at the $n$ locations where the squeezing occurs, becoming more negative with increasing $|\eta|$. Subplots \textbf{(c)} and \textbf{(l)} cover a larger area than the rest to fully showcase all of the relevant phase-space structures.}
    \label{fig:kl states}
\end{figure*}

A closed-form solution for the states satisfying the recursion relation in Eq. \eqref{eq:recursion relation} can be obtained for any choice of $k$ and $l$, but it becomes impractical for large $l$. Choosing the nonzero initial coefficient to be $\Psi_0$ and setting the next $l-1$ coefficients to zero, the resulting state can be written as
\begin{widetext}
\eq{
\ket{\Psi}\propto& \sum_{j=0}^\infty \ket{jn}\eta^j \sqrt{\left(jn\right)!}\prod_{i=0}^j\frac{(in+l)!}{(in+n)!}
=
\sum_{j=0}^\infty \ket{jn}\eta^j \sqrt{\left(jn\right)!}\frac{1}{\left(jn\right)!^{(n)} %
\left[j\left(n-1\right)\right]!^{(n)}\cdots
\left[j\left(k+1\right)\right]!^{(n)}
},
\label{eq:full state recursion}
}
\end{widetext} where $j!^{(n)}=j(j-n)(j-2n)\cdots$ is the $n$th factorial. One may use the property $(jn)!^{(n)}=j!n^n$ to recover the standard definition of single-mode squeezed states. The most noteworthy part of Eq. \eqref{eq:full state recursion} is that it only involves coefficients that differ by $n$. This property holds true regardless of the initial coefficient chosen to be nonzero, implying that all of the resultant states are unchanged via rotations by $2\pi/n$ in phase space, just like the higher-order cat states. The present states look like they have been squeezed from $n$ directions, which is another manner in which they generalize squeezed states.
We depict an exemplary array of generalized states satisfying Eq. \eqref{eq:creation annihilation eigenvalue kl} in Fig. \ref{fig:kl states}. 

The eigenvalue equation of Eq. \eqref{eq:creation annihilation eigenvalue kl} with $k=l$ and $\eta=1$  was recently investigated by Ref. \cite{PereverzevBittner2021}. There, the resultant states were deemed generalized coherent states because they were generated by the generalized squeezing operators of the form
\eq{
\hat{\mathcal{S}}(r,\varphi)=\exp\left(r\frac{\eu^{-\iu\varphi}\ha^k-\eu^{\iu\varphi}\had^k}{2}\right)\approx 1+r\frac{\eu^{-\iu\varphi}\ha^k-\eu^{\iu\varphi}\had^k}{2}
} in the $r\ll 1$ limit. Such states could thus be generated by optically pumping a crystal with a weak $k$th order nonlinearity. The states we find here broadly generalize Ref. \cite{PereverzevBittner2021}'s generalization to allow for $k\neq l$ and $\eta\neq 1$ in Eq. \eqref{eq:creation annihilation eigenvalue kl}.

The generalized squeezed states that we discuss here maximize the present notion of quantumness. Moreover, they are maximally sensitive to estimating arbitrary phase-space displacements for $n=k+l>2$, because they satisfy $\expct{\ha}=\Var(\ha)=0$ \cite{Goldbergetal2020extremal}, thereby extremizing another notion of quantumness. This shows the power of inequalities such as Eq. \eqref{eq:CS-squeeze-high} for determining the quantumness of a given quantum state.

\section{Discussion}
Optical devices that cause two input modes to interact are generally useful for creating entangled output modes. Since many such devices generate the least entanglement when the input states are coherent states, we defined a notion of quantumness as the amount of entanglement a state will generate with such an optical device. 

One common device for entanglement generation is a beam splitter. We showed that the unique type of state that generates the most entanglement at a beam splitter comprises a pair of single-mode squeezed states with equal squeezing, so long as the two squeezed states and the beam splitter obey a particular phase relationship. When the two squeezed states obey the opposite phase relationship, they generate no entanglement. Beyond being a peculiar result for our notion of quantumness, this could help define new notions: squeezed states maximize the amount of entanglement generated when maximized over all beam splitter phases, squeezed states maximize the difference between the amounts of entanglement generated by the optimal and worst beam splitter phases, and squeezed states may have their amounts of entanglement generated be the most sensitive to the beam splitter phase for some quantifier of sensitivity. All of these results follow because squeezed states uniquely satisfy the eigenvalue equation of Eq. \eqref{eq:creation annihilation eigenvalue}.

The entanglement generated by beam splitters is more useful in some circumstances than others. For example, the entanglement generated by a single photon impinging on a beam splitter can only be taken advantage of if one can distinguish between the vacuum and a single photon, giving rise to Hanbury-Brown and Twiss effects. Squeezed input states can lead to entanglement that is useful for phase sensing, which is considered to be a scenario where no entanglement needs to be provided to arrive at a quantum advantage \cite{Braunetal2018}. Other input states, such as Fock states in one mode and the vacuum in the other, provide no such advantage for phase sensing. We have thus focused exclusively on the ability of an input state to generate entanglement without considering its usefulness for a particular task.

Another common device for entanglement generation is a two-mode squeezer. With such a device, all input states generate at least \textit{some} entanglement. Again, as with beam splitters, a pair of single-mode squeezed states can generate both the most and the least entanglement, depending on the phase relationship between the squeezed states and the two-mode squeezer.
This helped us realize the general principle that the amount of entanglement generated by an optical device is highly dependent on the phase relationships between the input states and the device, which could again lead to a new notion of quantumness as the sensitivity of the amount of entanglement generated to changes in the phase of the device.

There are many circumstances in which the entanglement generated by a two-mode squeezer is useful. In fact, in most of those circumstances, nothing more than a vacuum-state input is necessary to provide some quantum advantage.
This distinction is apparent in resource theories of optical nonclassicality, in which beam splitters are considered to be free operations but two-mode squeezers are not \cite{Ferrarietal2020arxiv}. 
The ability of a pair of single-mode squeezed states to generate even more entanglement with a two-mode squeezer could lead to further enhancements in all of these tasks that use two-mode-squeezed-vacuum states.

These results are reminiscent of continuous-variable quantum key distribution \cite{GrosshansGrangier2002}, in which the best collective attack performed by an eavesdropper uses Gaussian transformations such as beam splitters and squeezers \cite{GrosshansCerf2004,Navascuesetal2006,GarciaPatronCerf2006}. Gaussian attacks are optimal because Gaussian states extremize strongly superadditive continuous functions that are invariant under local unitaries \cite{Wolfetal2006}. However, the scenario is different here: local unitaries \textit{do} affect the entanglement that a state generates at an optical device, so it does not immediately follow that Gaussian states such as coherent and squeezed-vacuum states will extremize the entanglement generated. There may therefore be deeper reasons as to why Gaussian states are extremal in these variegated scenarios.

These results are also directly applicable to coherent perfect absorption of quantum light. In Refs. \cite{HardalWubs2019,Vetlugin2021}, for example, it was shown that equally squeezed states input to a setup that classically leads to coherent perfect absorption of light are absorbed or not absorbed and become entangled or not entangled depending on the same phase relationships studied here. This further showcases the power of phase relationships in optical interactions and the peculiarity of squeezed states in these contexts.

Investigating the mathematics underlying this entanglement generation problem led us to consider states saturating the inequalities of Eqs. \eqref{eq:CS-cat-high} and \eqref{eq:CS-squeeze-high} as extremizing some notion of quantumness. The first inequality led to cat states and compass states, which have already proven to be useful in quantum information tasks such as metrology and error correction. The second led to a generalization of squeezed states, reminiscent of the so-called intelligent states \cite{Trifonov1994,Trifonov1997}, as those satisfying the eigenvalue equation of Eq. \eqref{eq:creation annihilation eigenvalue kl}. Such an equation dictates that losing $k$ photons from a state is in some sense equivalent to gaining $l$ photons, which could potentially have applications in quantum error correction. These states have remarkable phase-space properties as depicted in Fig. \ref{fig:kl states} that, at the very least, make them useful for sensing phase-space displacements in arbitrary directions, which could be applicable to tasks such as force sensing \cite{Dalvitetal2006}. There already exist schemes and proofs of principle for generating the cat and compass states and a scheme for creating some of the generalized squeezed states with $k=l$ and $\eta=1$. Future work could investigate schemes for creating states that satisfy Eq. \eqref{eq:creation annihilation eigenvalue kl} for arbitrary $k$ and $l$.

The ability of a state to generate entanglement at an optical device underlies a deep notion of quantumness and is intimately tied to many tasks in quantum information. We have restricted our attention to the entanglement between two modes with continuous-variable (i.e., Heisenberg-Weyl) phase spaces; it would be intriguing to study these phenomena in other phase spaces to better understand entanglement between and among spin systems, light, many-body systems, and beyond.

\begin{acknowledgments}
	The authors thank Aephraim Steinberg and Barry Sanders for insightful comments.
	AZG acknowledges funding from the Natural Sciences and Engineering Research Council of Canada, the Walter C. Sumner Foundation, and Cray Inc.
	\end{acknowledgments}

\onecolumngrid

\begin{appendix}
	\section{Entanglement generated by beam splitters, two-mode squeezers, and many other devices}
	\label{app:linear entropy general unitary}
	Beam splitters are represented by unitary operators $\hat{U}=\exp\left[-\iu\theta (\eu^{-\iu\psi}\had\hb+\eu^{\iu\psi}\ha\hbd)/2\right]$ [the extra phase $\phi$ in Eq. \eqref{eq:beam splitter matrix} does not affect the final entanglement properties]. We can generalize these linear operators to other, potentially nonlinear unitary operators, represented by 
	\eq{
		{\hat{\mathcal{U}}}	= \exp\left(r \frac{\eu^{-\iu\psi}\hoad\hob- \eu^{\iu\psi}\hoa\hobd}{2}\right),\quad r\geq 0.
		\label{eq:general unitary}
	} This form includes beam splitters, with $\hoa=\ha$, $\hob=\hb$, $r=\theta$, and the replacement $\psi\leftrightarrow \psi+\pi/2$; it also includes the single-mode squeezing operators from Eq. \eqref{eq:single-mode squeezing operator}, two-mode squeezing operators with $\hoa=\had$ and $\hob=\hb$, and more. We can investigate the amount of entanglement generated by these operators in the limit of small $r$, in order to find the states that are the most and least poised to generate entanglement with optical devices represented by unitaries of the form of Eq. \eqref{eq:general unitary}. In what follows, we make the assumption that $\hoa$ and $\hob$ act on different systems such that they and their Hermitian adjoints commute, precluding Eq. \eqref{eq:general unitary} from describing single-mode squeezing.
	
	An initially separable state $\ket{\Phi_{\mathrm{sep}}^{(a,b)}}=\ket{\Psi^{(a)}}\ket{\Psi^{(b)}}$ evolves under Eq. \eqref{eq:general unitary} to
	\eq{
		\ket{\Phi^\prime}&={\hat{\mathcal{U}}}\ket{\Phi_{\mathrm{sep}}^{(a,b)}}\\
		&= \left(1+r\frac{\eu^{-\iu\psi}\hoad\hob- \eu^{\iu\psi}\hoa\hobd}{2}+r^2\frac{\eu^{-2\iu\psi}\hoad^2\hob^2+\eu^{2\iu\psi}\hoa^2\hobd^2-\hoad\hoa\hob\hobd-\hoa\hoad\hobd\hob}{8}\right)\ket{\Phi_{\mathrm{sep}}^{(a,b)}}+\mathcal{O}(r^3).
	} Defining $\hat{\sigma}=\ket{\Psi^{(a)}}\bra{\Psi^{(a)}}$, we can trace out mode $b$ from $\ket{\Phi^\prime}$ to find $\hat{\rho}_a\equiv\Tr_b(\ket{\Phi^\prime}\bra{\Phi^\prime})$:
\eq{
\hat{\rho}_a=&\hat{\sigma}+r\frac{ \eu^{-\iu\psi}\expct{\hob}\left(\hoad \hat{\sigma}-\hat{\sigma}\hoad\right) + \eu^{\iu\psi}\expct{\hobd}\left(\hat{\sigma}\hoa -\hoa \hat{\sigma}\right) }{2}\\
&+r^2\frac{ \expct{\hobd\hob} \hoad\hat{\sigma}\hoa+ \expct{\hob\hobd} \hoa\hat{\sigma}\hoad -\eu^{-2\iu\psi}\expct{\hob^2}\hoad\hat{\sigma}\hoad -\eu^{2\iu\psi}\expct{\hobd^2}\hoa\hat{\sigma}\hoa  }{4} \\
&+r^2\frac{ \eu^{-2\iu\psi}\expct{\hob^2}\hoad^2\hat{\sigma} + \eu^{2\iu\psi}\expct{\hobd^2}\hoa^2\hat{\sigma}
+\eu^{-2\iu\psi}\expct{\hob^2}\hat{\sigma}\hoad^2 + \eu^{2\iu\psi}\expct{\hobd^2}\hat{\sigma}\hoa^2
 }{8}\\
&-r^2\frac{\expct{\hob\hobd}\hoad\hoa\hat{\sigma}+\expct{\hobd\hob}\hoa\hoad\hat{\sigma}
	+\expct{\hob\hobd}\hat{\sigma}\hoad\hoa+\expct{\hobd\hob}\hat{\sigma}\hoa\hoad}{8}+\mathcal{O}(r^3).
} Squaring this quantity, taking the trace, and collecting like terms yields
\eq{
\Tr(\hat{\rho}_a^2)=&1-\frac{r^2}{2}\left(\expct{\hoa\hoad}-\left|\expct{\hoad}\right|^2\right)\left(\expct{\hobd\hob}-\left|\expct{\hob}\right|^2\right)-\frac{r^2}{2}\left(\expct{\hoad\hoa}-\left|\expct{\hoad}\right|^2\right)\left(\expct{\hob\hobd}-\left|\expct{\hob}\right|^2\right)\\
&+r^2\RE\left[\eu^{-2\iu\psi}\Var\left(\hoad\right)\Var\left(\hob\right)\right]+\mathcal{O}(r^3),
} where all of the expectation values are taken with respect to the initial state $\ket{\Phi_{\mathrm{sep}}^{(a,b)}}$, or, equivalently, with respect to $\ket{\Psi^{(a)}}$ and $\ket{\Psi^{(b)}}$. The linear entropy thus becomes
\eq{
\frac{2H_{{\hat{\mathcal{U}}}}\left(\Phi_{\mathrm{sep}}^{(a,b)}\right)}{r^2}=&\left(\expct{\hoa\hoad}-\left|\expct{\hoad}\right|^2\right)\left(\expct{\hobd\hob}-\left|\expct{\hob}\right|^2\right)
+
\left(\expct{\hoad\hoa}-\left|\expct{\hoa}\right|^2\right)\left(\expct{\hob\hobd}-\left|\expct{\hob}\right|^2\right)\\
&-2\RE\left[\eu^{-2\iu\psi}\Var\left(\hoad\right)\Var\left(\hob\right)\right]+\mathcal{O}(r).
\label{eq:general linear entropy}
} We repeatedly use this result in the main text.
We further note that, for arbitrary $\hoa$ and $\hob$, a sufficient condition for generating the minimum amount of entanglement given by Eq. \eqref{eq:general linear entropy} is to be an eigenstate of $\hoa$ and $\hob$ or to be an eigenstate of $\hoad$ and $\hobd$. 

For beam splitters, with $\hoa=\ha$ and $\hob=\hb$ each satisfying bosonic commutation relations, we simply use that $\ha\had=\had\ha+1$ and $\hb\hbd=\hbd\hb+1$ to achieve the final result of Eq.
\eqref{eq:linear entropy second order}.
For two-mode squeezing operators, with $\hoa=\had$ and $\hob=\hb$, this commutation relation yields a result slightly different from Eq. \eqref{eq:linear entropy second order}:
\eq{
\frac{2H_{\mathrm{two-mode\,squeezing}}\left(\Phi_{\mathrm{sep}}^{(a,b)}\right)}{r^2}=2AB+A+B+1-2\RE\left[\eu^{-2\iu\psi}\Var\left(\ha\right)\Var\left(\hb\right)\right]+\mathcal{O}(r),
} where we have again defined $A=\expct{\had\ha}-\left|\expct{\ha}\right|^2\geq 0$ and $B=\expct{\hbd\hb}-\left|\expct{\hb}\right|^2\geq 0$. We discuss this in further detail in the main text.

More general observations also follow. Just like for beam splitters and two-mode squeezers, the entanglement generated is only maximized when $\expct{\hoa}=\expct{\hob}=0$. In that case, we can express the linear entropy as
\eq{
\frac{2H_{\hat{\mathcal{U}}}\left(\Phi_{\mathrm{sep}}^{(a,b)}\right)}{r^2}=
\expct{\hoa\hoad}\expct{\hobd\hob}+
\expct{\hoad\hoa}\expct{\hob\hobd}
-2\cos\Theta\left|\expct{\hob^2}\right|\left|\expct{\hoa^2}\right|+\mathcal{O}(r)
,
} where we have defined the phase
\eq{
\Theta=\arg\left[\Var\left(\hob\right)\right]-\arg\left[\Var\left(\hoad\right)\right]-2\psi.
} 
If we find states saturating the inequalities
\eq{
\left|\expct{\hoa^2}\right| \leq \sqrt{\expct{\hoa\hoad}\expct{\hoad\hoa}}\quad \mathrm{and} \quad 
\left|\expct{\hob^2}\right| \leq \sqrt{\expct{\hobd\hob}\expct{\hob\hobd}},
} which generalize Eq. \eqref{eq:CS-squeeze} and are saturated by states satisfying the Eq. \eqref{eq:creation annihilation eigenvalue}-like eigenvalue equations
\eq{
\hoa\ket{\left(\Phi_{\mathrm{sep}}^{(a,b)}\right)}\propto \hoad\ket{\left(\Phi_{\mathrm{sep}}^{(a,b)}\right)}\quad\mathrm{and}\quad\hob\ket{\left(\Phi_{\mathrm{sep}}^{(a,b)}\right)}\propto \hobd\ket{\left(\Phi_{\mathrm{sep}}^{(a,b)}\right)},
\label{eq:eigenvalue general operators}
} we can express the linear entropy in the compact form
\eq{
\frac{2H_{\hat{\mathcal{U}}}\left(\Phi_{\mathrm{sep}}^{(a,b)}\right)}{r^2}=
\left|\sqrt{\expct{\hoa\hoad}\expct{\hobd\hob}}-\eu^{\iu\Theta}
\sqrt{\expct{\hoad\hoa}\expct{\hob\hobd}}\right|^2+\mathcal{O}(r).
}
Just like for beam splitters and two-mode squeezers, the entanglement generated depends quite strongly on the phase relationship between the two input states and the optical element, characterized by $\Theta$. The commutation relations $\left[\hoa,\hoad\right]$ and $\left[\hob,\hobd\right]$ are important to the final result, and the ability to find states that satisfy Eq. \eqref{eq:eigenvalue general operators} for generalized operators such as $\hoa=\had^k$ is crucial to finding states that extremize the entanglement generated by arbitrary optical elements.

\end{appendix}

\end{document}